\documentclass[aps,pra,twocolumn]{revtex4}

\usepackage[english]{babel}
\usepackage{graphicx}
\usepackage{amssymb}
\usepackage{amsmath, amsthm, amssymb}
\usepackage{amsthm}

\newcommand{\ctg}{\mathop{\rm ctg}\nolimits}
\newcommand{\arctg}{\mathop{\rm arctg}\nolimits}
\newcommand{\Ei}{\mathop{\rm Ei}\nolimits}

\begin{document}

\title{Low-dimensional weakly interacting Bose gases: non-universal equations of state}

\author{G.E.~Astrakharchik$^1$, J.~Boronat$^1$, I.L.~Kurbakov$^2$, Yu.E.~Lozovik$^2$, F.Mazzanti$^1$}

\affiliation{$^1$ Departament de F\'{\i}sica i Enginyeria Nuclear, Campus Nord B4-B5, Universitat Polit\`ecnica de Catalunya, E-08034 Barcelona, Spain}
\affiliation{$^2$ Institute of Spectroscopy, 142190 Troitsk, Moscow region, Russia}

\date{\today}

\begin{abstract}
The zero-temperature equation of state is analyzed in low-dimensional bosonic systems. In the dilute regime the equation of state is universal in terms of the gas parameter, {\it i.e.} it is the same for different potentials with the same value of the $s$-wave scattering length. Series expansions of the universal equation of state are reported for one- and two- dimensional systems. We propose to use the concept of energy-dependent $s$-wave scattering length for obtaining estimations of non-universal terms in the energy expansion. We test this approach by making a comparison to exactly solvable one-dimensional problems and find that the generated terms have the correct structure. The applicability to two-dimensional systems is analyzed by comparing with results of Monte Carlo simulations. The prediction for the non-universal behavior is qualitatively correct and the densities, at which the deviations from the universal equation of state become visible, are estimated properly. Finally, the possibility of observing the non-universal terms in experiments with trapped gases is also discussed.
\end{abstract}


\maketitle

\section{Introduction}

Understanding the properties of rarefied quantum systems is a fundamental question that has been addressed in a large number of works. This problem was extensively studied in the 50s-60s when significant development of mathematical formalism (perturbative methods, Feynman diagrams, diagonalization techniques, etc., see, for example, \cite{LLIII,LLIX}) permitted to obtain important results and brought a lot of interest to dilute quantum systems. Some important results were as well obtained in low-dimensional systems \cite{Girardeau60,Lieb63}, which at that moment were rather mathematical toys with reduced applicability in the real world. The situation changed radically with the realization of Bose-Einstein condensation in dilute gases\cite{Anderson95,Davis95}. Having an excellent experimental control over the geometry of the cloud it was possible to create essentially pure quantum gases in the dilute regime and to probe the system properties.
The experimental advances in the field with the realization of very anisotropic traps stimulated further the interest in dilute low dimensional gases (see, for example, \cite{Gorlitz01,Schreck01,Greiner01,Moritz03,Tolra04,Stoferle04}).

In the ultradilute limit the interparticle potential can be described by one parameter, namely the $s$-wave scattering length $a$, and the ground state properties of a gas are governed by the gas parameter $na^D$, where $n$ is the particle density and $D$ stays for the dimensionality. As the density is increased details of the interaction potential become important. Such a  non-universal regime has been thoroughly studied in three-dimensional geometries, where the universal terms are known \cite{Lee57,Lee57b}. Low energy corrections coming from the specific interaction potential can be described by the effective range $r_0$. Corrections to the ground-state energy, excitation spectrum and condensate fraction can be obtained (see, for example, \cite{Beliaev58,Brueckner57} and more recent works \cite{Braaten01,Bulgac02}).
It was shown that for two body problem the inclusion of an energy-dependent pseudopotential improves significantly upon the use of an energy-independent pseudopotential \cite{Blume02b}. Also, the concept of momentum dependent scattering length is very useful for estimation of the interaction for a Rydberg atom where it allows to take into account the effect of the Coulomb potential of nucleus, see for example Ref~\cite{Bendkowsky09}.
Unfortunately, much less is known in low-dimensional systems. Indeed, only recently the universal terms of the 2D equation of state have been correctly derived\cite{Cherny01,Mora03,Pricoupenko04} and checked numerically \cite{Astrakharchik09a}. In the present study we address the problem of non-universal corrections in low-dimensional systems.

The rest of the article is organized as follows. In Sec.~\ref{sec:a(k)} we discuss the origins of the universal behavior and study the two-body scattering problem and propose a simple way to obtain non-universal corrections. In Sec.~\ref{sec:One-dimensional systems} the equation of state of some exactly solvable one-dimensional models are analyzed. Some properties of two-dimensional systems are addressed in Sec.~\ref{sec:Two-dimensional systems}. We start with an overview of the literature in Sec.~\ref{sec:2D overview}. In Sec.~\ref{sec:universal terms} we discuss the expansion of the universal equation of state and provide some physical insight on the origins of the beyond mean-field (BMF) terms. The knowledge of the expansion of the universal equation of state permits us to investigate the non-universal equation of state as it comes from the method proposed in Sec.~\ref{sec:a(k)} and confront that with numerical results. Section~\ref{sec:non-universal terms} is devoted to the study of non-universal effects in the $s$-wave scattering problem and in the many-body equation of state. In Sec.~\ref{sec:discussion} we discuss the possibility of experimental observation of non-universal effects in trapped cold gases. The feasibility of reaching an ultradilute two-dimensional regime is also discussed. Finally, the main conclusions are drawn in Sec.~\ref{sec:conclusions}.

\section{Universal and non-universal terms \label{sec:a(k)}}

In dilute systems the probability of three-body collisions is highly reduced leaving the two-body scattering the most important physical process. In this process two particles scatter each other with a relative momentum $k$. The two-body scattering problem is described by the Schr\"odinger equation
\begin{eqnarray}
-\frac{\hbar^2}{2\mu}\Delta\psi(r)+V_{int}(r)\psi(r) = \frac{\hbar^2k^2}{2\mu}\psi(r),
\label{psi}
\end{eqnarray}
where $\mu$ is the reduced mass. If the interaction potential $V_{int}(r)$ is short-ranged, its exact shape is not important at low density and the relevant quantity of the scattering solution $\psi(r)$ is the phase $\delta(k)$ at distances larger than the range of the potential. For small scattering energies the phase can be expanded in terms of the momentum $k$. In a 3D system this leads to
\begin{eqnarray}
k\ctg \delta(k) = -\frac{1}{a_0}+\frac{1}{2}k^2 r_0+...,
\label{a1}
\end{eqnarray}
where $a_0$ is the $s$-wave scattering length and $r_0$ is the effective range. If the scattering momentum is very small the only relevant parameter is the $s$-wave scattering length $a_0$ and all potentials having the same value of $a_0$ will behave similarly. This limit is known as {\it universal regime}. The relevant length scales are then $a_0$ and the interparticle distance. It is expected that the many-body ground-state energy can be expressed in terms of the gas parameter $na_0^D$, where $D$ denotes the dimensionality of the problem.

For example, the low density energy per particle of a homogeneous weakly-interacting Bose gas in 3D at zero temperature is given by
\begin{eqnarray}
\frac{E_{3D}}{N}=\frac{2\pi\hbar^2na_0}{m}\left(1+\frac{32}{15\pi}\sqrt{16\pi na^3_0}+... \right) \label{E3D}
\end{eqnarray}
with the leading term linear in the density being the mean-field Gross-Pitaevskii contribution\cite{Bogoliubov47} and quantum fluctuations contributing to the subleading $n^{3/2}$ Lee-Huang-Yang correction\cite{Lee57,Lee57b}. The next term scales like $n^{2}$, but it is no longer universal\cite{Beliaev58,Brueckner57} and depends on the explicit choice of the interaction potential.

It is possible to recast the definition (\ref{a1}) of the scattering length $a_0$ in a different form, namely, as the position of the node of the analytic continuation of the scattering solution from distances much larger than the range of the potential in the zero-energy scattering limit. Indeed, in 3D, in the limit of very low-scattering energy the phase reads $\delta(k)=-ka_0$ and the scattering solution becomes $\sin(kr+\delta)/r\to k(r-a_0)/r$, which has a node at $r=a_0$. The advantage of the alternative definition is that it is well suited also to low-dimensional problems.

\newtheorem{mydef}{Definition}
We generalize the last definition to finite values of the scattering energy.
\begin{mydef} The generalized scattering length $a(k)$ is the position of the node of the analytical continuation of the large distance $r\to\infty$ two-body scattering solution $\psi(r)$ at the scattering energy $\hbar^2k^2/2\mu$. If there are several nodes, the position of the closest node to $r=0$ is considered.\label{def:a(k)}\end{mydef}
In this way the $s$-wave scattering length $a(k)$ depends on the scattering momentum and fulfills the condition $\lim_{k\to 0} a(k)=a_0$. An example how the $s$-wave scattering length changes with the type of the potential is shown in Fig.~\ref{Fig:wf} for several characteristic interactions in one dimension. The figure shows the asymptotic continuation of zero energy scattering solution. We will show in the next sections that the inclusion of the finite-momentum corrections improves the description of the energy and allows us to estimate the term of the expansion where the non-universal behavior appears.


At this point it is important to understand the relation between the effective range and the $s$-wave scattering length in the description of non-universal effects. The effective-range theory is well established in three-dimensional systems (see, for example, textbook \cite{LLIII}). The effective range is then defined from the expansion of the phase shift in terms of the scattering momentum, see Eq.~(\ref{a1}). The constant term defines the $s$-wave scattering length $a_0$, and the effective range $r_0$ corresponds to taking into account dependence on $k^2$. Instead, the energy dependent $s$-wave scattering length $a(k)$ includes in addition all higher order momenta, i.e. $k^2$, $k^4$, $k^6$,... More importantly the concept of $a(k)$ can be applied to low-dimensional systems, where the non-universal terms in the equation of state are not generally known. In our approach it is enough to know the dependence on $a_0$ of the universal equation of state and the non-universal terms will be automatically generated.

\section{One-dimensional systems \label{sec:One-dimensional systems}}

One peculiarity of the one-dimensional world is that several many-body models can be solved {\em exactly} (with short- \cite{Girardeau60,Lieb63} and long- \cite{Sutherland71} range interactions), in the sense that the exact ground state can be written either explicitly\cite{Girardeau60,Sutherland71} or can be easily obtained as the solution of a system of integral equations \cite{Lieb63}. This allows us to test the proposed approach of using an energy dependent scattering length by comparing to the exactly known results.

The ground-state energy of a Bose gas with a repulsive $\delta$-pseudopotential interaction (Lieb-Liniger model) can be obtained by solving Bethe {\it ansatz} equations. The expansion of the energy in the mean-field regime\cite{Lieb63} has a structure similar to that of the three-dimensional case (\ref{E3D}):
\begin{eqnarray}
\frac{E_{1D}}{N} = \frac{1}{2}g_{1D}n_{1D} \left( 1-\frac{4\sqrt2}{3\pi}(n|a_{1D}|)^{-1/2}+... \right) \label{E1D}
\end{eqnarray}
where $g_{1D}=-2\hbar^2/(ma_{1D})>0$ is the one-dimensional coupling constant. Indeed, the leading term in Eq.~(\ref{E1D}) is the same as it would come out from the mean-field Gross-Pitaevskii equation, while the subleading term is the same as obtained from Bogoliubov theory. In passing by we note that such a coincidence is not obvious {\it a priori}, as both the Gross-Pitaevskii and Bogoliubov theories assume that all or a large fraction of particles are in the condensate. Instead, strictly speaking, Bose-Einstein condensation in homogeneous one-dimensional system is absent\cite{Hohenberg67}.

The reason why the theories based on the presence of a Bose condensate produce correct results for energetic properties can be understood by following the similar arguments used in the renormalization group approach (see, e.g. Ref.~\cite{Popov83}). The main contribution to the energy comes from short distances. At short distances the phase coherence may be present even in the absence of the Bose-Einstein condensation. Therefore, on this length scale it is possible to apply the perturbative theories that are based on the assumptions of a macroscopic occupation of the condensate. Coherence at finite distances larger than the interparticle distance is sufficient for MF and Bogoliubov theories to yield correct result for the ground-state energy.  In particular, such theories successfully describe one-dimensional systems at zero temperature (such as Lieb-Liniger gas in the regime of weak correlations) and two-dimensional dilute Bose gas at finite temperature, none of which has true Bose-Einstein condensation. A mathematical way to resolve the paradox and to prove the validity of the Bogoliubov result in one-dimensional systems is to use space discretization and to introduce the concept of a quasi-condensate\cite{Castin03}.

\begin{figure}
\begin{center}
\includegraphics[angle=-90,width=0.5\columnwidth]{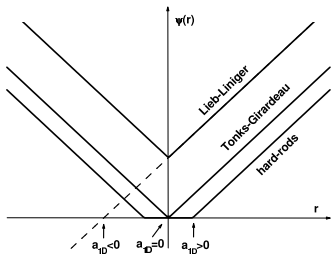}
\caption{Solid lines: typical two-body scattering solutions $\psi(r)$ at zero energy for Lieb-Liniger (upper curve), Tonks-Girardeau (middle curve) and Hard-Rod (lower curve) Hamiltonians. Dashed line: analytic continuation of the scattering solution for the Lieb-Liniger model. Arrows point the positions of the nodes.} \label{Fig:wf}
\end{center}
\end{figure}

Contrary to three- and two- dimensional systems, here the mean-field regime means high densities $n_{1D}a_{1D}\gg 1$. This precludes us from using the concept of energy-dependent scattering length in the MF regime, as the energy of an incident particle would be huge, see Eq.~(\ref{E1D}), and so would be the deviations of $a(k)$ from $a_0$. Thus, the mean-field regime is no longer universal (contrary to what happens in 3D and 2D systems), as the energy and correlation functions are very different for the $\delta$-pseudopotential \cite{Lieb63,Astrakharchik03,Astrakharchik06b}, the Calogero-Sutherland $1/z^2$ potential \cite{Sutherland71,Astrakharchik06c} and  the dipolar $1/|z|^3$ interaction \cite{Arkhipov05}. Instead, in the regime of strong quantum correlations, $n|a_{1D}|\ll 1$, the energy and correlation functions of all those models (essentially, for any repulsive interaction potential) approach the same {\em universal} limit referred as Tonks-Girardeau\cite{Girardeau60} regime (see also Fig.~\ref{Fig:wf}).

This is a peculiarity of the one-dimensional world that the dilute regime, $n_{1D}|a_{1D}|\ll 1$ corresponds not to a mean-field limit, but rather to a regime where quantum fluctuations are dominant. The energy in this limit is given by the energy of an ideal Fermi gas $E/N=\pi^2\hbar^2n_{1D}^2/(6m)$ and the wave function of strongly interacting bosons can be mapped onto a wave function of non-interacting fermions\cite{Girardeau60,Mazzanti08,Mazzanti08b}. For instance, the energy of a gas of hard rods of size $a_{1D}>0$ is obtained from the energy of an ideal Fermi gas by taking into account the excluded volume\cite{Girardeau60}: $n_{1D}\to N/(L-Na_{1D})$. We expand this expression in terms of the one-dimensional gas parameter $\tilde n = n_{1D}a_{1D}$ at small densities $n\ll 1$ and get
\begin{eqnarray}
\frac{E_{HR}}{N} 
=\frac{\pi^2\hbar^2n_{1D}^2}{6m}(1+2\tilde n+3\tilde n^2+4\tilde n^3+...).
\label{EHR}
\end{eqnarray}

It is worth mentioning that the beyond-mean-field terms in three-dimensional systems were first obtained for a hard-sphere gas by Lee, Huang, Yang\cite{Lee57,Lee57b} and afterwards were shown to be  universal\cite{Brueckner57,Beliaev58,Lieb63}. Starting from the expansion for a one-dimensional analogue for hard-spheres, Eq.~(\ref{EHR}), we will calculate the first non-universal corrections for a different potential. We chose a $\delta$-pseudopotential, as its exact groundstate energy is known and thus we can test our approach. The solution of the scattering problem (\ref{psi}) with $V_{int}(r) = g_{1D}\delta(r)$ can be readily written $\psi(r) \propto \sin (k|r|-\arctg ka_{1D})$. The energy-dependent $s$-wave scattering length can be explicitly expressed as a function of the momentum and the leading correction to $a_0$ is quadratic in momentum
\begin{eqnarray}
a(k) = \frac{\arctg ka_{1D}}{k}
= a_{1D} - \frac{1}{3} k^2a^3_{1D} + ...
\label{ak}
\end{eqnarray}
The substitution of (\ref{ak}) into (\ref{EHR}) for a characteristic value of the energy $\hbar^2k^2/m \propto \pi^2\hbar^2n_{1D}^2/(6m)$ allows us to estimate the first correction due to non-universality:
\begin{eqnarray}
\frac{E_{LL}^{appr.}}{N} =
\frac{\pi^2\hbar^2n_{1D}^2}{6m}
\left(\!1+2\tilde n+3\tilde n^2+ \left[4\!-\!\frac{\pi^2}{9}\right]\tilde n^3+...
\right)
\label{ELLapprox}
\end{eqnarray}
A possible concern about the validity of the obtained result is that expansion (\ref{EHR}) is done for $n_{1D}a_{1D} >0$, while expansion (\ref{ELLapprox}) is used to describe a region where $n_{1D}a_{1D}<0$, with a different sign of the $s$-wave scattering length. We argue that the universal equation of state is smooth as a function of the one-dimensional gas parameter $n_{1D}a_{1D}$. This is supported by the apparent similarities between the hard-rod gas and the gas-like state of the attractive $\delta$-pseudopotential (``super-Tonks-Girardeau'' system)\cite{Astrakharchik05b}. We also note that the Bethe {\it ansatz} solution for two-component attractive and repulsive fermions is continuous (compare results of Refs.~\cite{Gaudin67,Yang67,Krivnov75}).

The result can be compared to the exact predictions for the Lieb-Liniger model based on the Bethe {\it ansatz} technique. The exact result can be obtained by solving the integral equations recursively (details are given in Appendix~\ref{derivationELL}) and reads
\begin{eqnarray}
\frac{E_{LL}}{N} =\frac{\pi^2\hbar^2n_{1D}^2}{6m}
\left(\!1+2\tilde n+3\tilde n^2+ \left[4\!-\!\frac{14\pi^2}{15}\right]\tilde n^3+...\right)
\label{ELL}
\end{eqnarray}

By comparing the exact results for the hard-rod gas, Eq.~(\ref{EHR}), exact results for the $\delta$-pseudopotential gas, Eq.~(\ref{ELL}), and the approximate result (\ref{ELLapprox}) obtained by the proposed method we conclude that:

\begin{itemize}
\item The order of the expansion in which the $\delta$-pseudopotential and hard rod energies differ is predicted correctly
\item The expansion (\ref{ELL}) contains same rational terms as the expansion (\ref{EHR}), while in addition it has irrational terms (here multiples of $\pi^2$). The use of energy-dependent scattering length permits to guess correctly the structure of the potential-dependent correction.
\end{itemize}

We find that the first three terms of the expansion are the same for considered potentials. The physical meaning of such terms is that particles behave as if they were ideal fermions in the box of size $L-Na_{1D}$. Indeed, this interpretation explicitly applies to the hard-rod gas, where the excluded volume correction is negative, as $a_{1D}>0$. For a negative scattering length the ``excluded volume'' correction changes its sign and becomes positive $L\to L+N|a_{1D}|$. In Fig.~\ref{Fig:wf} we present the characteristic behavior of the one-body scattering solution at low energy for three short-ranged potentials. The Tonks-Girardeau potential corresponds to zero-range infinitely strong repulsion. This places the node of the wave function at the origin and according to definition~\ref{def:a(k)}, the value of the $s$-wave scattering length is zero $a_{1D}=0$. For a hard-rod interaction potential the position of the first node is positive and thus $a_{1D}>0$. The slope of $\psi(r)$ is determined by the scattering momentum (refer to Eq.~(\ref{ak}) and the discussion above it), so for a similar scattering energy the only relevant difference in the wave function corresponding to different interactions is just a shift in abscissas. Thus, two Tonks-Girardeau particles separated by a distance $r$ and two hard-rod particles separated by a distance $r-a_{1D}$ ``feel'' each other in the same way. The only differences appear at very small distances of the order $r\approx a_{1D}$. In a similar way the scattering solution for a Lieb-Liniger Hamiltonian can be ``adjusted'' to match the Tonks-Girardeau solution by the change $r\to r-a_{1D}$ (mind that $a_{1D}<0$ in this case). This makes natural that the ``excluded volume'' correction is encountered for different potentials and it can change its sign. This was first noted in Ref.~\cite{Astrakharchik06a}.

\section{Two-dimensional systems \label{sec:Two-dimensional systems}}
\subsection{Overview of the equation of state\label{sec:2D overview}}

A series of our previous works have been devoted to the study of the equation of state of dilute two-dimensional Bose gases \cite{Mazzanti05,Astrakharchik07a,Astrakharchik07,Lozovik07,Lozovik08,Astrakharchik08,Astrakharchik09a}. A number of interaction potentials (dipolar, Yukawa, etc.) were considered in a wide range of densities. There it was demonstrated that the dipoles crystalize at large densities. The density of the quantum phase transition turned out to be extremely large $na_0^2\approx 2900$ \cite{Astrakharchik07a,buchler07,Mora07}. This shows that dipolar interaction potential is rather ``soft'' compared to the hard-core potentials, which are expected to crystallize at values of the gas parameter smaller than unity (for example, $na^2_0 = 0.33(2)$ in the case of hard-disks \cite{Xing90}). The equation of state of a dilute gas was obtained both in the universal and non-universal regimes from Monte Carlo calculations for dipoles \cite{Mazzanti05,Astrakharchik08}, hard- and soft- disks \cite{Pilati05}. A peculiarity of the two-dimensional systems is that the Gross-Pitaevskii equation has a limited applicability even in very dilute systems \cite{Astrakharchik07} due to the logarithmic dependence on the gas parameter rather than on powers of it, like in three- and one- dimensional systems. This means that it is extremely difficult to study numerically the universal equation of state. Densities as low as $na^2 \approx 10^{-100}$ had to be reached in calculations \cite{Astrakharchik09a} in order to check numerically the low-density expansion of the universal equation of state. It turns out that in order to describe correctly the beyond mean-field effects, several terms have to be summed even at densities as low as $na_0^2\approx 10^{-10}$ since the series comes out in terms of the slow converging logarithm function $\ln na^2_0$, which at such densities is of the order of the next (constant) term. Historically it turned out to be very difficult to obtain the correct expression for this term, see Ref.~\cite{Astrakharchik09a} for a summary of different results. Only recently the correct expression was obtained \cite{Cherny01,Mora03,Pricoupenko04,Mora09}.

Once the structure of the universal terms is well established, we are ready to test the concept of an energy-dependent $s$-wave scattering length.


\subsection{Universal terms \label{sec:universal terms}}

In this section we study the equation of state of 2D Bose gases in the universal regime, {\it i.e.} where the interaction potential can be described by one parameter, namely the $s$-wave scattering length, and all properties of the gas are fully defined by the gas parameter $na^2$. Without going through a rigorous derivation of the equation of state, which would be an extremely tedious calculation, we provide some simple ideas that give insight into the relevant physics involved in the equation of state.

In a weakly interacting system of any dimensionality the leading contribution to the energy comes from the mean-field theory. Assuming that the density is low enough, it does not matter what the exact shape of the short-range interaction potential is, and a simple $\delta$-pseudopotental can be used. In this way the real interaction potential can be replaced by a zero-range one, such that it imposes a correct zero boundary condition to the scattering state
\begin{equation}
V(r) = g\delta(r)\times(regularization)
\label{pseudopotetnial}
\end{equation}
The regularization operator is needed to make the $\delta$-function description compatible with a generic $1/r$ (or $1/\sqrt{r}$) divergence in a 3D (or 2D) geometry, although this is not important for our considerations. The substitution of (\ref{pseudopotetnial}) into the expression of the interaction energy written in first quantization simplifies the double integration
$E = 1/2 \int d{\bf r}_1\int d{\bf r}_2 \hat\Psi^\dagger({\bf r}_1)\hat\Psi^\dagger({\bf r}_2)V(|{\bf r}_1-{\bf r}_2|)
\hat\Psi({\bf r}_2)\hat\Psi({\bf r}_2)
= g/2 \int\!d{\bf r}\;\hat\Psi^\dagger({\bf r})\hat\Psi^\dagger({\bf r})\hat\Psi({\bf r})\hat\Psi({\bf r})$. Treating the field operator $\hat \Psi({\bf r})$ as a classical field and substituting it with the particle density $\sqrt{n}$ one obtains the mean-field expression for the energy
\begin{equation}
\frac{E}{N} = \frac{1}{2}gn
\label{MF}
\end{equation}

It is easy to see from Eq.~(\ref{pseudopotetnial}) that the coupling constant has dimensionality of [$E\times L^D$] and it has to be expressed in terms of the parameters of the scattering problem, which are $\hbar, m$ and $a$. In the three-dimensional case the considerations of units leads to a combination proportional to the $s$-wave scattering length $g_{3D}\propto \hbar^2 a_{3D}/m$. Indeed, the exact expression is $g_{3D} = 4\pi \hbar^2 a_{3D}/m$. In a one-dimensional system the correct units are obtained in a combination which is inversely proportional to the $s$-wave scattering length $g_{1D}\propto \hbar^2/(ma_{1D})$. This agrees with the exact result $g_{1D}=-2\hbar^2/(ma_{1D})$. The two-dimensional case is special in the sense that combinations having the proper units can be obtained without involving the $s$-wave scattering length $g_{2D}\propto \hbar^2/m$. The dependence on $a$ can come only in a combination with the scattering momentum $k$, which in a homogeneous system is related to the density. The exact result \cite{Schick71,Lieb01} indeed has the anticipated structure $g_{2D} = 4\pi \hbar^2/(m|\ln na^2_{2D}|)$. As explained, the corresponding mean-field term can be split into a part independent of $a$
\begin{equation}
\frac{E^{MF,(0)}}{N} = \frac{2\pi\hbar^2n}{m}
\label{MF0}
\end{equation}
and a part that depends on $a$
\begin{equation}
\frac{E^{MF,(1)}}{N} = \frac{2\pi\hbar^2n}{m}\frac{1}{|\ln na^2|}
\label{MF1}
\end{equation}
The momentum dependence of the 2D coupling constant makes perturbative theory very involved. Moreover, the dependence of the coupling constant on the $s$-wave scattering length is very weak as $a$ enters in a logarithm. Already at the mean-field level one sees that the usual relation between the chemical potential $\mu^{MF} = g n$ and the energy per particle $E^{MF}/N = 1/2\; g n$ written as $E/N = 1/2\;\mu$ is valid only to the leading term. Indeed, direct integration of the chemical potential $E(N) = \int_0^N \mu(N') dN'$ leads to a different result
\begin{eqnarray}
\frac{E^{\text{MF}}}{N} = \frac{2\pi\hbar^2\Gamma(0,2|\ln n a^2|)}{m na^4}
\label{EMF}
\end{eqnarray}
where $\Gamma(a,x) = \int_{x}^\infty t^{a-1} e^{-t} dt$ is the incomplete gamma function. The energy expansion in the dilute regime $na^2\to 0$ can be obtained by doing the integration by parts or from the large argument expansion of $\Gamma(0,x)$. One finds that this generates a number of terms including a contribution to the BMF constant
\begin{eqnarray}
\frac{E^{\text{MF}}}{N}
= \frac{2\pi n\hbar^2/m}{|\ln na^2| + 1/2 - 1/(4|\ln na^2|) + ...}
\label{EMFexp}
\end{eqnarray}
The difference contain terms that are logarithmically small which, anyway, exceeds accuracy of the mean-field theory, so that at the MF level there is no legitimate reason to prefer one expression over the other. On the other hand terms of this order are important when one studies beyond MF terms.

The most important beyond mean-field terms were obtained by V. N. Popov\cite{Popov72} in 1972 (see also his book \cite{Popov83}). He obtained a recursive expression relating the chemical potential $\mu$ and the density $n$ for a given value of the inverse temperature $\beta$
\begin{eqnarray}
n = \frac{m\mu}{4\pi\hbar^2}\left(\ln\frac{\varepsilon_0}{\mu}-1\right)
-\int \frac{\hbar^2k^2}{2m\varepsilon(k)}\frac{1}{e^{\beta\varepsilon(k)}-1}
\frac{d^2k}{(2\pi)^2}
\label{Popov}
\end{eqnarray}
where $\varepsilon^2(k) = (\hbar^2k^2/2m)^2+\hbar^2k^2\mu/m$ is the Bogoliubov spectrum and $\varepsilon_0$ is of the order of $\hbar^2/mr_0^2$, with $r_0$ is the range of the interaction potential. We write the last relation introducing an unknown coefficient of proportionality $C_1$ such that $\varepsilon_0 = C_1 \hbar^2/ma^2$.

At zero temperature quasiparticle excitations are absent and the expression simplifies. By solving Eq.~(\ref{Popov}) iteratively one obtains the following expression for the chemical potential
\begin{eqnarray}
\mu^{popov} = \frac{4\pi\hbar^2 n/m}{|\ln na^2|+\ln |\ln {na^2}|-\ln 4\pi+\ln C_1-1...}
\label{mupopov}
\end{eqnarray}

Lozovik and Yudson used in 1978 diagrammatic techniques to find a recursive relation which relates the chemical potential and the density \cite{Lozovik78}
\begin{eqnarray}
\mu=\frac{4\pi\hbar^2n_0/m}{|\ln(\mu ma_0^2/\hbar^2)|+O(1)}.
\label{Elozovik}
\end{eqnarray}
Solving recursively Eq.~(\ref{Elozovik}) [see also \cite{Popov83} and Appendix B in \cite{Schick71}] one generates the first BMF term $\ln|\ln na^2|$ and the second BMF term proportional to $\ln \pi$.

It should be noted that while in a three-dimensional system Bogoliubov approximation works in a dilute regime, in the two-dimensional case even in a very rarified gas the Bogoliubov theory becomes inapplicable\cite{Lozovik78}.

Furthermore, the condensate fraction has a logarithmic dependence on the gas parameter $n_0/n = 1-1/|\ln na^2|$ \cite{Schick71, Popov71}. The difference between the total density $n$ and the condensate density $n_0$ in Eq.~(\ref{MF}) leads to an additional contribution in the second BMF (constant) term.

Another contribution to BMF terms should contain Euler's $\gamma$ constant. The calculation of the chemical potential summing the ladder diagrams relates $\mu$ to the scattering amplitude\cite{Beliaev58}, which itself contains the $\gamma$ constant in the expansion\cite{Hines78}.

Summarizing, one expects to find the following types of BMF corrections:
\begin{itemize}
\item A first BMF term of the form $\ln |\ln na^2|$
\item A contribution to the second BMF (constant) term proportional to $\ln \pi$
\item A contribution to the second BMF (constant) term proportional to $\gamma$
\item A contribution to the second BMF (constant) term proportional to a constant of the order of $1$
\end{itemize}

Historically it took a long time to obtain correctly BMF expansions at low densities (for a literature review refer to \cite{Astrakharchik09a}). The double logarithm term can be obtained from the iterative relation (\ref{Elozovik}) and it is present in the majority of theories. Unfortunately, this term alone is not sufficient to describe the universal regime and the calculation of the all the contributions to the second BMF term was a challenging task. We note that the corresponding problem in 3D was solved in the 1950s \cite{Lee57b} and the 1D problem in 1960s \cite{Girardeau60,Lieb63}.

The universal equation of state for the chemical potential should read then
\begin{eqnarray}
\mu= \frac{4\pi\hbar^2n/m}{
     |\ln na^2|
+ \ln|\ln na^2| + C^{\mu}_1
+\frac{\ln|\ln na^2| +C^{\mu}_2}{|\ln na^2|}+...
},
\label{muan}
\end{eqnarray}
Notice that this expression is compatible both with Eq.~(\ref{mupopov}) and the result of iterating Eq.~(\ref{Elozovik}) for $\mu$. The second BMF term was recently obtained analytically \cite{Cherny01,Mora03,Pricoupenko04} as $C^{\mu}_1 = -\ln\pi -2\gamma -1 = - 3.30...$ and its value was confirmed numerically in Ref.~\cite{Astrakharchik09a}. The subsequent constant was derived a short time ago in Ref.~\cite{Mora09} with its value given by $C^{\mu}_2 = -0.751$. In the following we will use a value obtained from a fit to Monte Carlo data $C^{\mu}_2 = -0.3(1)$ \cite{Astrakharchik09a}.

The expansion of the energy per particle takes then a form similar to (\ref{muan})
\begin{eqnarray}
\frac{E}{N} = \frac{2\pi\hbar^2n/m}{
     |\ln na^2|
+ \ln|\ln na^2| + C^{E}_1
+\frac{\ln|\ln na^2| + C^{E}_2}{|\ln na^2|}+...
},
\label{Ean}
\label{Efit}
\end{eqnarray}
with the coefficients related as $C^{E}_1 = C^{\mu}_1+1/2 = - 2.80...$ and $C^{E}_2 = C^{\mu}_2+1/4$ (equals to $-0.05(10)$ from the numerical fit).

\subsection{Non-universal terms \label{sec:non-universal terms}}

The specific details of an interaction potential become important when the density is large, so that the equations of state is no longer universal. In the regime of high densities one parameter, namely, the zero-energy $s$-wave scattering length $a$ is no longer sufficient to describe the system properties. In Section~\ref{sec:a(k)} we have formulated our proposal using an energy dependent $s$-wave scattering length. This allows us to generate non-universal terms in an energy expansion and also to understand analytically at which densities deviations from the universal law appear. This can be applied at densities for which the universal equation of state is known. As the reference equation of state we take (\ref{Efit}). In this section we test our proposal for three different potentials, such as hard disks, soft disks and dipoles.

In the case of hard disks, the interaction has only one length scale, namely, the size of the disk. As a result the energy dependence is trivial $a_{HD}(k) = a_0$. The absence of additional length scales means that out of all interaction potentials the hard core potential has an equation of state that is the most similar to the universal one.

In the case of soft disks, corrections due to the finite scattering energy are important at typical densities $na^{2}\gtrsim 10^{-3}$ \cite{Pilati05}. The first correction due to the finite value of the scattering energy is quadratic in momentum, as shown in Appendix~\ref{appendix:aSD}. The explicit expression for $a(k)$ is given by formula (\ref{aSD1}) and it reduces to $a_{SD}(k)=a_{SD}(0)(1-5.53454k^{2}+...)$ for the choice of soft disk parameters as in Ref.~\cite{Pilati05}.

\begin{figure}
\begin{center}
\includegraphics[angle=-90,width=\columnwidth]{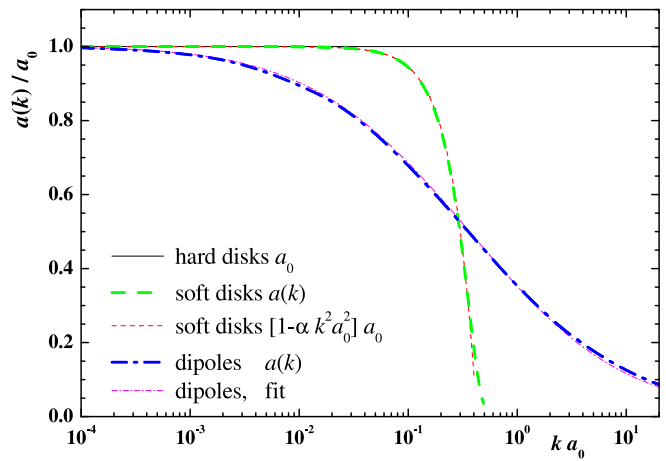}
\caption{(Color online) Finite-energy $s$-wave scattering length as a function of momentum $k$ of the incident particle for different interaction potentials. All quantities are measured in units of $a_0$. Solid line, hard disks $a(k)=a_0$; dashed lines, soft disks: thick line, numerical solution as the node of (\ref{fSD}), thin line, analytical expansion $a(k)/a_0 = 1 -5.53454 k^2a_0^2$ as comes from (\ref{aSD1}) using the parameters of soft disks taken from Ref.~\cite{Pilati05}; dash-dotted lines, dipoles: thick line, numerical solution, thin line, fit (\ref{add}).}
\label{Fig:a}
\end{center}
\end{figure}

The dipolar interaction potential decays slowly and deviations from the universal equation of state appear much earlier. In a very dilute system, $na^{2}\lesssim 10^{-7}$, the dipole-dipole scattering length is well approximated by its value at zero scattering momentum, $a_{dd}(0)=e^{2\gamma}r_d=3.17222...r_d$, where $r_d$ is a characteristic lengthscale for dipole-dipole interaction potential \cite{Astrakharchik07a}. We solve the $s$-wave scattering problem numerically and find that the following fit describes well the numerical data for the value of the $s$-wave scattering length at low energies
\begin{eqnarray}
\frac{a_{dd}(k r_d)}{a_{dd}(0)}\!=\!e^{-\!\exp\{0.441082\!+\!0.31414\ln k r_d\!-\!0.0275752\ln^{2}k r_d\}}
\label{add}
\end{eqnarray}

In order to find non-universal corrections to the energy, according to the proposed scheme, we substitute the gas parameter $na^2_0$ in the universal expansion (\ref{Efit}) with $na^2(k)$. Within the level of accuracy of interest, it is sufficient to use the mean-field expression for the scattering momentum $k^{2}\propto 2mE/N\hbar^2=4\pi n/|\ln na^{2}|$.

In the case of soft disks this leads to the substitution $\ln na^{2}\rightarrow \ln na_{0}^{2}+2\ln (1-\alpha k^2a_0^2)$. The logarithm can be expanded as $\ln(1-\varepsilon)=\varepsilon +...$, leading to non-universal corrections of the order of $2\alpha k^2a_0^2\propto 8\pi \alpha na^2/|\ln na^{2}|$. The resulting equation of state for soft disks than reads
\begin{widetext}
\begin{eqnarray}
\frac{E}{N} = \frac{2\pi\hbar^2n/m}{
     |\ln na^2|
+ \ln|\ln na^2| -\ln\pi -2\gamma -1/2
+[\ln|\ln na^2| -\ln\pi -2\gamma +2.0(1) +1/4  + 8\pi \alpha na^2] / |\ln na^2|
}
\label{EanSD}
\end{eqnarray}
\end{widetext}

\begin{figure}
\begin{center}
\includegraphics[angle=-90,width=\columnwidth]{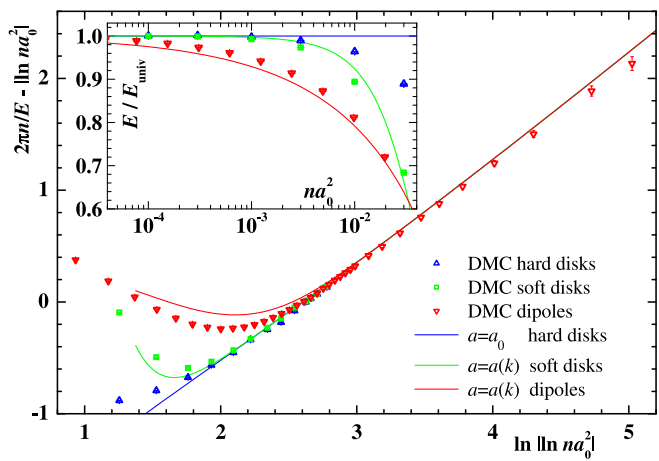}
\caption{(Color online) Energy per particle, analysis of non-universal beyond MF corrections. Main figure, Beyond MF terms in the energy as a function of the double logarithm of the gas parameter. Symbols, DMC results: up triangles, hard disks, squares, soft disks, down triangles dipoles.
Corresponding lines: hard disks, Eq.~(\ref{Ean}), soft disks, Eq.~(\ref{EanSD}), dipoles Eq.~(\ref{Ean}) with $a(k)$ as in Eq.~(\ref{add}).
Inset, energy per particle $E/N$ in units of ``universal'' equation of state, Eq.~(\ref{Ean}), as a function of the gas parameter $na^2$.}
\label{Fig:Eas}
\end{center}
\end{figure}

The obtained analytical expressions for the equation of state are confronted with the results of Monte Carlo simulations. Figure~\ref{Fig:Eas} shows the beyond MF energy as a function of the double logarithm of the density for different interaction potentials (compare to Fig.~1 in Ref.~\cite{Astrakharchik09a}). As anticipated, the beyond MF terms have the most simple dependence for hard disk potential (since the $s$-wave scattering length dependence is flat, see Fig.~\ref{Fig:a}) and it is the best one described by the ``universal'' equation of state (\ref{Ean}). The region where the description of the energy is universal shrinks in the case of soft disks and diminishes further for dipoles (compare to the dependence of the corresponding $a(k)$, Fig.~\ref{Fig:a}). We find that the analytical description we obtain for the non-universal behavior works rather well. In particular, the density, at which deviations from the universal law start to be visible, is predicted correctly by our approach. The analytical formula (\ref{EanSD}) provides not only a good qualitative description, but even the quantitative agreement is good. From Fig.~\ref{Fig:Eas} it might seem that the description is better for soft disks compared to dipoles, but in reality the description for both potentials is expected to have a similar level of accuracy. In order to check that we solved self-consistently the equations $E = E(n,u,a); a = a(E); u = u(n,a)$, 
(where $u$ is a dimensionless in-medium scattering amplitude \cite{Cherny01}, see also Appendix~\ref{Appendix:B})
thus obtaining a different expression, which have the same significative perturbation terms, but differ in higher order terms which are outside of the accuracy of our approach. The self-consistent solution improves coincidence for dipoles, but also changes the predictions for the soft disks introducing deviations similar to the ones of dipoles in Fig.~\ref{Fig:Eas}.

\section{Discussion \label{sec:discussion}}

Recent progress in techniques of cooling and confinement permits to realize extremely dilute gases in the regime of quantum degeneracy, thus providing a very advanced tool for studying properties of weakly interacting gases. The $s$-wave scattering length can be controlled by use of Feshbach resonance and can be set to, essentially, any desired value by choosing an appropriate magnetic field. Many features of the equation of state can be inferred from measuring energetic properties, such as release energies in time of flight experiments. Also the size of the cloud and the density profile are related to the equation of state. The most precise technique for the moment is the accurate measurement of the frequencies of collective oscillations. This method was successfully used to study beyond MF terms in the equation of state of two-component Fermi gases in the BCS-BEC crossover\cite{Altmeyer07}.

In previous sections we have investigated the properties of low-dimensional weakly interacting Bose gases as a function of the one- and two-dimensional gas parameters $n_{1D}a_{1D}$ and $n_{2D}a_{2D}^2$, respectively. The low-dimensional system can be realized in experiments by strongly squeezing the gas in one or two directions. Assuming that the trapping is harmonic with frequency $\omega$ the condition of being in a low-dimensional regime is that the oscillator levels should not be excited neither by the energy per particle nor by the temperature $E/N, k_BT\ll\hbar\omega$.

In a one-dimensional system a relation between the three-dimensional $a_{3D}$ and the one-dimensional $a_{1D}$ $s$-wave scattering lengths was found in Ref.~\cite{Olshanii98} assuming harmonic radial confinement with oscillator length $a_{ho}$. The relation has a resonant behavior when $a_{3D}$ is of the same order as $a_{ho}$ due to the contribution of virtual excitations of the levels of transverse confinement. The one-dimensional coupling constant $g_{1D} = -2\hbar^2/(ma_{1D})$ is expressed as \cite{Olshanii98}
\begin{eqnarray}
g_{1D} = \frac{2\hbar^2}{m a_{ho}^2}\frac{1}{1-1.0326\,a_{3D}/a_{ho}}
\label{resonance:1D}
\end{eqnarray}
In particular, at the top of an Olshanii resonance, the one-dimensional $s$-wave scattering vanishes and $a_{1D}=0$, making $g_{1D}\to\infty$. This corresponds to the Tonks-Girardeau limit. Close to the resonance $a_{1D}$ is small and expansions like (\ref{EHR},\ref{ELLapprox}) are applicable.

In a similar way to Eq.~(\ref{resonance:1D}), the coupling constant in a quasi-two-dimensional system has a resonant structure \cite{Petrov00b}
\begin{eqnarray}
g_{2D} = \frac{4\pi\hbar^2}{m}\frac{1}{|\ln(2\pi |\mu|m a_{ho}^2/\hbar^2)| + \sqrt{2\pi}\, a_{ho}/a_{3D}}
\label{resonance:2D}
\end{eqnarray}
and describes a competition between ``purely two-dimensional'' logarithmic term and a mean-field Gross-Pitaevksii term $g^{GP}_{Q2D} = 2\sqrt{2\pi}\hbar^2/m$, which can be obtained from the Gross-Pitaevskii energy functional assuming a Gaussian profile in the tight direction of the confinement. The results from Section~\ref{sec:Two-dimensional systems} apply to a purely two-dimensional system, when the logarithmic term in Eq.~(\ref{resonance:2D}) is dominant, {\it i.e.} when $a_{ho}\ll a_{3D}$. We have explained that the mean-field (here, in a ``purely'' two-dimensional sense) regime is achieved when the double logarithm of the two-dimensional parameter is large $\ln|\ln na^2|\gg 1$ which leads to extremely rarefied densities, such as $na^2\ll 10^{-862}$. Fortunately, the nature provides a way to get such small effective densities. Indeed, Eq.~(\ref{resonance:2D}) can be rewritten introducing the second term of the denominator under the logarithm. The resulting expression can be interpreted in the sense of a ``purely two-dimensional'' system, but with a rescaled effective density $n^\star \propto exp\{ -\sqrt{2\pi} a_{ho}/a_{3D}\} n$. Here the effect of large $a_{ho}/a_{3D}$ ratios is exponentially amplified.

Expressions for the (quasi)low-dimensional coupling constants (\ref{resonance:1D},\ref{resonance:2D}) were obtained from the analytic solution of the two-body scattering problems in the presence of a tight harmonic confinement. The existence of one-dimensional resonance in a many-body system was later confirmed in numerical simulations \cite{Astrakharchik02b}. A similar two-dimensional study is more involved as the expression of the coupling constant depends on the chemical potential and we are not aware of such studies.

\begin{figure}
\begin{center}
\includegraphics[angle=-90,width=\columnwidth]{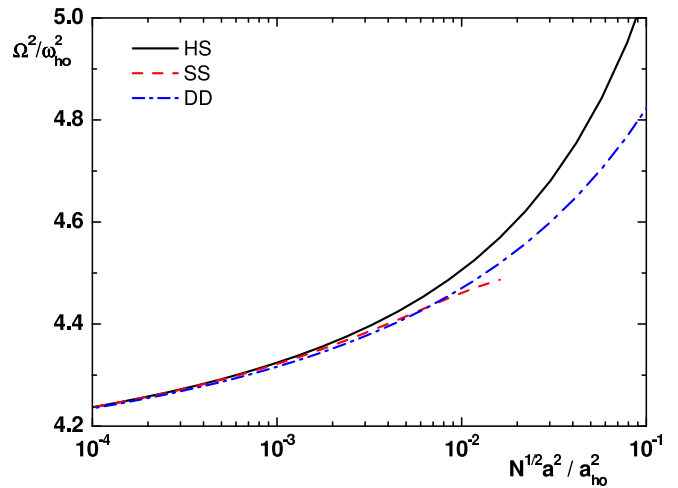}
\caption{(Color online)
Comparison of the lowest breathing mode frequency, $\Omega^2$ as a function of the coupling strength $N^{1/2}a^2/a_{ho}^2$ for different interaction potentials. Solid line, hard disks, Eq.~(\ref{Ean}); dashed line, soft disks, Eq.~(\ref{EanSD}); dash-dotted line, dipoles, Eq.~(\ref{Ean}) with $a(k)$ as in Eq.~(\ref{add}).}
\label{Fig:freq:pot}
\end{center}
\end{figure}
The energetic properties of trapped gases can be accessed by observing the frequencies of collective oscillations. By displacing the center of the trap it is possible to generate oscillations which depend only on the frequency of the trapping potential. Instead, a sudden change in the frequency of the trap causes ``breathing'' oscillations, for which the frequency depends on the compressibility of the gas, which, in turn, is related to the equation of state. In Fig.~\ref{Fig:freq:pot} we show predictions for different interaction potentials in two-dimensional systems. Out of all considered model interactions, the hard core potential shows the strongest dependence. The soft disk and dipolar potentials have softer dependencies.

\section{Conclusions \label{sec:conclusions}}

To conclude, we have studied the energetic properties of dilute low-dimensional interacting Bose gases at zero temperature. In the regime of ultralow densities the equation of state is described by only one dimensionless parameter, the gas parameter $na^D$. The universal equation of state in three- and one- dimensional systems dates back to 1960s
\cite{Lee57b,Girardeau60,Lieb63}. The beyond mean-field terms of the two-dimensional equation of state were obtained recently in Refs.~\cite{Cherny01,Mora03,Pricoupenko04} and their correctness was verified in Monte Carlo calculations \cite{Astrakharchik09a}. When the density is increased, the details of the interaction potential become important and deviations from the universal behavior are observed.

We propose to use an energy-dependent $s$-wave scattering length to describe the non-universal behavior. This method permits to generate non-universal terms in the equation of state. The advantage of the proposed approach is that it sufficient to known the dependence on the $s$-wave scattering length for this method to be applicable. This permits us to use it in low dimensional systems, where the equation of state in term of non-universal are not well established.
We test this approach on one-dimensional systems, where a direct comparison to exactly solvable models is done, and on two-dimensional systems, where numerical results for different interaction potentials (hard disks, soft disks, dipoles  are used. We find that the typical density at which the non-universal terms become important is correctly estimated. For one-dimensional systems, in the cases when the energy expansion can be obtained exactly, we show that the structure of the potential-dependent terms is predicted correctly. Finally we point out that non-universal terms can be studied experimentally by observing frequencies of collective oscillations.


\section{Acknowledgements}

The work was partially supported by (Spain) Grant No. FIS2005-04181, Generalitat de Catalunya Grant No. 2005SGR-00779 and RFBR. G.E.A. acknowledges post doctoral fellowship by MEC (Spain).

\appendix
\section{Exact equation of state of weakly-interacting one-dimensional Bose gas with $\delta$-pseudopotential interactions.\label{derivationELL}}

The ground-state energy of a Lieb-Liniger gas can be found exactly using the Bethe {\it ansatz} approach. The energy as a function of the gas parameter is obtained implicitly by solving the following system of integral equations\cite{Lieb63}:
\begin{eqnarray}
\label{LL:e}
e(\gamma)&=& \frac{\gamma^3}{\lambda^3}\int\limits_{-1}^1 k^2 \rho(k)\;dk\\
\label{LL:gamma}
\gamma&=&\lambda\left / \int\limits_{-1}^1\rho(k)\;dk\right.\\
\label{LL:rho} \rho(k)&=&\frac{1}{2\pi}+\int\limits_{-1}^1\frac{2\lambda\rho(\varkappa)}{\lambda^2+(k-\varkappa)^2} \;\frac{d\varkappa}{2\pi},
\end{eqnarray}
where $\gamma=-2/n_{1D}a_{1D}$.

It is possible to obtain explicit expressions for the energy in terms of the gas parameter in the limits of small and large gas parameter as a series expansion. We solve the system of Eqs.~(\ref{LL:e}-\ref{LL:rho}) iteratively in the regime $|na|\ll 1$. This is done by starting from $\rho^{(0)}=1/(2\pi)$ and substituting it into the r.h.s. of Eq.~(\ref{LL:rho}) to get $\rho^{(1)}$. The obtained expression is used for the next iteration and so on.

We provide an explicit expression for the ground-state energy close to the Tonks-Girardeau regime:
\begin{eqnarray}
\frac{E}{N}\!\!\! &=&\!\!\! \frac{\pi^2\hbar^2n^2}{6m}\!\left[ \sum\limits_{l=0}^\infty (l\!+\!1)\left(\!-\frac{2}{\gamma}\!\right)^l\!+\!\frac{32\pi^2}{15\gamma^3}
\!+\!O(\gamma^{-4}\!)
\right]
\label{ELL:TGexpansion}
\end{eqnarray}

The first two terms: $E/N = \frac{\pi^2\hbar^2n^2}{6m} (1 -4/\gamma)$ were obtained in the original work of Lieb and Liniger \cite{Lieb63} (see also \cite{Haldane81,Brand04}).

We also rewrite expression (\ref{ELL:TGexpansion}) in terms of the gas parameter $\tilde n = n_{1D}a_{1D}$ as
\begin{eqnarray}
\frac{E}{N} &=& \frac{\pi^2\hbar^2n^2}{6m} \left[\sum\limits_{l=0}^\infty (l+1)\tilde n^l
- \frac{4\pi^2\tilde n^3}{15}
+O(\tilde n^4)
\right]
\end{eqnarray}
The ``excluded volume'' contribution is intentionally separated from the non-universal part.

\section{Equation of state of a two-dimensional Bose gas from Cherny and Shanenko theory.\label{derivationEOS2D}\label{Appendix:B}}

We note that the derivation of the equation of state of a weakly interacting Bose gas, proposed by Cherny and Shanenko \cite{Cherny01}, can be used to obtain an explicit expression of the energy as a function of the density $n$:
\begin{eqnarray}
\frac{E}{N} = \frac{2\pi\hbar^2n(u)}{m}
\left[u+\frac{u^2}{2}+2u^2e^{2/u}\Ei\left(-\frac{2}{u}\right)\right],\\
n(u)a^2=\frac{\exp\{-2\gamma-1/u\}}{\pi u},
\end{eqnarray}
here $u$ (dimensionless in-medium scattering amplitude) defines the parametrical dependence of the energy on the density, $\Ei(x)=-\int_x^\infty e^{-t}/t\;dt$ is exponential integral function, and $\gamma$ is Euler's constant.

\section{Finite-energy scattering problem for a soft disk potential in 2D.\label{appendix:aSD}}

In order to find the two-dimensional $s$-wave scattering length of the soft-disk potential in 2D we have to solve the two-body scattering problem. The positive energy Scr\"odinger equation for two particles of equal mass reads
\begin{eqnarray}
-\frac{\hbar^2}{m}\Delta f(r) + V(r) f(r) = \frac{\hbar^2k^2}{m}f(r),
\end{eqnarray}
where $k$ is the relative momentum. We consider the soft disk interaction potential and look for a spherically symmetric solution. The interaction potential is defined by the range of the potential $R_0$ and the height of the soft disk by
\begin{eqnarray}
V(r) =
\left\{
{\begin{array}{ll}
\hbar^2\varkappa^2/m,& |r| \le R_0\\
0, & |r|>R_0
\end{array}}
\right.
\label{VHD}
.
\end{eqnarray}

In the inner region, $r<R_0$, we use a solution that is regular at the origin,
\begin{eqnarray}
f(r) = I_0(r\sqrt{\varkappa^2-k^2}), \quad  |r|<R_0,
\end{eqnarray}
where $I_0(x)$ is the modified Bessel function of the first kind. The normalization constant is not important for the present considerations. In the outer region the solution is simply a two-dimensional plane wave
\begin{eqnarray}
f(r) = C_1 J_0(k r) + C_2Y_0(k r), \quad  |r|>R_0,
\end{eqnarray}
where $J_0(x)$ and $Y_0(x)$ are Bessel functions of the first and second kind, respectively. The coefficients $C_1$ and $C_2$ are obtained from the continuity condition for $f(r)$ and $f'(r)$ at the edge of the soft disk $r=R_0$. This gives the following solution in the outer region, $r>R_0$.
\begin{eqnarray}
\nonumber
f(r)&=&\frac{\pi R_0}{2} \left\{\kappa I_1(\kappa R_0) \left[J_0(kR_0)Y_0(kr)-J_0(kr)Y_0(kR_0)\right] \right . \\
&+&\left. k I_0(\kappa R_0)\left[J_1(kR_0)Y_0(kr)-J_0(kr)Y_1(kR_0)\right]\right\},
\label{fSD}
\end{eqnarray}
where $\kappa = \sqrt{\varkappa^2-k^2}$. The $s$-wave scattering length is the node of the function (\ref{fSD}) closest to the origin. We will consider the case of low densities, so that the incident particles is slow, $k\ll \varkappa$. Then $f(r)$ can be expanded in powers of $k$ and one has
\begin{eqnarray}
\nonumber
f(r) &=& I_0(\varkappa R_0) + \varkappa R_0 I_1(\varkappa R_0)\ln(r/R_0)\\
\nonumber
&+& \frac{k^2R_0^2}{4} \left[(1-r^2/R_0^2) I_0(\varkappa R_0)\right.\\ \nonumber
&-& \left. \frac{2 + \varkappa^2(R_0^2-r^2) - \varkappa^2(r^2+R_0^2)\ln(R_0/r)}{\varkappa R_0} I_1(\varkappa R_0)\right]\\
&+&O(k^4)
\label{fSD2}
\end{eqnarray}
The zero energy $s$-wave scattering length is found by setting $k=0$ in (\ref{fSD2}) and leads to
\begin{eqnarray}
a_0 = \exp\left\{-\frac{I_0(\varkappa R_0)}{\varkappa R_0 I_1(\varkappa R_0)}\right\}R_0
\label{aSD0}
\end{eqnarray}
Furthermore, one can set $r\to a_0$ in the second line of Eq.~(\ref{fSD2}) and find a correction to the position of the node
\begin{eqnarray}
a(k) = a_0 - \alpha k^2a_0^3 + O(k^4),\\
\alpha =
\frac{R_0}{4\varkappa a_0^2}\frac{I_0(\varkappa R_0)+I_2(\varkappa R_0)}{I_1(\varkappa R_0)}
-\frac{1}{4}\left(\frac{R_0^2}{a_0^2}-1\right)
\label{aSD1}
\end{eqnarray}
In order to test the accuracy of Eq.~(\ref{aSD1}) we also find numerically the nodes of the finite-energy scattering function (\ref{fSD}). A comparison of the exact result for the $s$-wave scattering length $a(k)$ to the expansion (\ref{aSD1}) is presented in Fig.~\ref{Fig:a}. We see that at the densities of interest, the obtained expansion works very well.

\end{document}